# Self-Organization on Unstable Vicinal Surfaces with Competing Interactions

V. Tonchev*, B. Ranguelov, D. Staneva
Institute of Physical Chemistry, Bulgarian Academy of Sciences,
1113 Sofia, Bulgaria

Long-ranged step-step attractions destabilize the vicinal crystal surfaces. Their competition with shorter-ranged step-step repulsions results in self-organized patterns. The exponent in the time-scaling of their characteristic size is influenced only by the range of the attractions but not by the range of the repulsions as we show based on precise numerical analysis of two different minimal models. Another anisotropy we identify is that the vicinal surface is not destabilized by shorter-ranged attractions.

PACS: 81.10.Aj, 68.35.Rh, 89.75.Da

The vicinal crystal surfaces are consequence of the discrete nature of matter on atomic scale – cutting through the crystal along a plane with arbitrary angle results in a set of equidistant monatomic steps separated by plain terraces. These 'nano-stairways' play important role in the contemporary vacuum technologies and focus significant efforts in the fundamental science as well since the steps mediate the processes of crystal growth in the so called *step flow growth mode*. Their motion could be unstable in different contexts leading to the formation of groups of steps (bunches) and extended terraces. These patterns are subject to further interest as templates on the self-organized route towards the synthesis of various nano-structures. Classical destabilizing mechanisms are the Ehrlich-Schwoebel effect and ad-atom electromigration and a complete theory of bunching of straight steps due to these is formulated in [1].

In this paper we follow the consequences of another cause of destabilization – the step-step attraction. The possibility for switching on step-step attraction when a metal vicinal surface is put into contact with an electrolyte is studied experimentally [2] and theoretically [3]. Step bunch formation breaks down the Gaussian distribution of the step-step distances into a double peaked one as observed on Ag(19, 19, 17)–vicinals [2]. The tendency to lower the surface tension, with two ingredients – step-step attraction and repulsion, through a phase separation into a 'hill and valley' structure is identified by theoretical considerations [3], the step-step attraction supposed to be of electrostatic origin.

In such a case the general step-step interaction energy $E$ of two steps placed a distance $d$ apart could be written in the form:

$$E = \frac{U_0}{d^n} - \frac{K_0}{d^p} \qquad (1)$$

$n$ ($p$) is the range of the repulsion (attraction) and $U_0$ ($K_0$) – it's magnitude. This type of interaction is reflected in the equation(s) of step motion in the LW2 model [4]:

$$\frac{dx_i}{dt} = U(2f_i - f_{i-1} - f_{i+1}) - K(2g_i - g_{i-1} - g_{i+1}) \qquad (2)$$

$x_i$ is the position of the $i$-th step in a (1+1)D step train, $U$ and $K$ are properly rescaled magnitudes of the interactions and $f_i$ is defined as:

$$f_i = \left(\frac{1}{\Delta x_i}\right)^{n+1} - \left(\frac{1}{\Delta x_{i+1}}\right)^{n+1} \qquad (3)$$

$\Delta x_i = x_i - x_{i-1}$ and $g_i$ is obtained from $f_i$ when changing $n$ with $p$. Only the distances between steps that are nearest neighbors are taken into account in the above equations. In parallel we use as an important reference a *toy* model - MM2, constructed *ad hoc* [5] as:

$$\frac{dx_i}{dt} = Uf_i - Kg_i \tag{4}$$

In MM2 the meaning of the parameters $U$ and $K$ is not the same as in LW2. In both equations $U$ ($K$) contains the vicinal distance $l$ raised to the power $n+1$ ($p+1$). In order to find the values of the parameters that make the vicinal surface unstable we carry out a simplest version of stability analysis - linear stability of the motion of a step in the step train. Note that the equidistant step train has zero velocity. Now we perturb only the position of $i$-th step with $\delta l$. This results in a non-zero step velocity $\delta V_i$. The step motion is unstable when $\delta V_i . \delta l > 0$ and after plugging into the equations of step motion the changed terrace widths $\Delta x_i = l + \delta l$ and $\Delta x_{i+1} = l - \delta l$ and keeping only terms that are linear in $\delta l$ one obtains the stability condition:

$$\left(\frac{n+1}{p+1}\frac{U}{K}\right)^{1/(n-p)} \equiv S\left(\frac{n+1}{p+1}\right)^{1/(n-p)} < l \tag{5}$$

same for both models. The scaling parameter $S$ will appear also in the numerical results for developed instability. This condition predicts two instability scenarios: (i) shorter-ranged repulsion $n>p$ with $U<K$ and (ii) shorter-ranged attraction $n<p$ with $U>K$; but our numerical study ruled out the second one – bunches form only when the repulsion is shorter-ranged. This finding confronts the paradigm in which the short-ranged attraction competes with long-ranged repulsion in the faceting on Si(113) vicinal surfaces.

We study further the models using a computational protocol combining numerical integration of the velocity equations (IV-th order Runge-Kutta procedure) and a self-consistent method to recognize the surface pattern using two monitoring schemes [6] as described briefly: Monitoring Scheme I (MS-I) - at every discrete advance of the integration is counted the number of step bunches, using the only natural definition – a distance between two adjacent steps is considered as *bunch distance* when it is smaller than the initial (vicinal) distance. Further is found the average bunch size $N$ as the sum of all steps in bunches divided by the number of bunches, etc.; Monitoring Scheme II (MS-II) - for every bunch size up to the maximal one that will be encountered during the calculation are cumulated quantities as the minimal step-step distance $l_{min}$, bunch width, etc., and finally these are averaged over the number of times they occur. The 'bunch width vs. bunch size' dependences from both monitoring schemes lie on a single curve thus proving their use to be self-consistent [6]. The time-scaling of the (average) number of steps in the bunch $N$ from MS – I and size-scaling of the (averaged) minimal step-step distance in the bunch $l_{min}$ from MS – II and complementary morphological information for the patterned surface are necessary to determine the model's universality class. Same self-consistent approach is recommended when treating experimental data.

The numerical integration starts with almost perfect vicinal surface with steps slightly deviated from their equilibrium positions in order to instigate the instability. Once started the process continues with step rearrangement into bigger and bigger groups (bunches) while the mass center of the step system remains intact since there is no growth or evaporation flux. Qualitatively LW2 generates the specific type of step bunches with constant slope of the crystal surface inside [5,7] the bunch of any size, observed also in experimental study of step bunching on TaC (910) complemented with Monte Carlo simulation which implements long-ranged attraction and shorter-ranged repulsion [8]. The same type of bunching was obtained from a similar to LW2 model with $p = 0$ and $n = 2$ [9]. The constant values of the 'minimal' step-step distance in the bunch, obtained with broad range of model parameters were used to obtain [4] the scaling relation:

$$l_{\min} = S \equiv \left(\frac{U}{K}\right)^{1/(n-p)} \qquad (6)$$

We checked this dependence for MM2 and the result is shown on Figure 1, identical to the one from LW2. Thus, the only way to distinguish between the two models remains to explore the time dependence of the number of steps in the bunch $N$ as could be anticipated from partial results obtained in [4,5].

In next plots, Figures 2, 3 and 4, we show the time dependence of the (averaged) number of steps in the bunch $N$ for different values of $n$ and $p$ and the obtained time-scaling exponent $\beta$. Clear is the asymmetry in the role of $n$ and $p$ - $\beta$ changes only with changing the range of attractions $p$ but not with changing the range of repulsions $n$. Based on a precise systematic study comprising much more values of $p$ we obtain the general dependence of $\beta$ on $p$:

$$\beta = \beta_0 (1 - qp) \qquad (7)$$

(LW2) $\quad \beta_0 = 1/5, q = 1/10 \qquad\qquad\qquad \beta_0 = 1/3, q = 1/5 \quad$ (MM2)

First we check this result in the context of the developed continuum approach [1, 10]. In these papers is proposed a generalized continuum equation for the time evolution of the surface height consisting of two terms - stabilizing and destabilizing one. We construct a new continuum equation following the same procedure of constructing MM2 and LW2 – the new destabilizing term is obtained from the repulsions term with inverting its sign and changing the exponent in it – $n$ for $p$. Analyzing the new continuum equation using a scaling approach as in [10] we obtain $\beta = 0.25$ independent of $p$ and $n$. Therefore, further theoretical efforts are needed to arrive at the proper continuum analogue of our model(s).

Our study provides an additional perspective on some previous results. In [7] was studied a model identical to MM2 with $n = 2$, $p = 1$. The exponent in the time-scaling of the terrace width was identified as 1/4 but it is actually a little bit larger - 0.271 (0.003) found from the digitized data and very close to the value 0.26(6) as follows from Eq.7. In experimental studies of faceting kinetics on Si(113) vicinal surfaces by Song et al. [11,12] where direct step-step attraction is supposed as driving the surface transformation different values of $\beta$ were found as function of the quench temperature. Although the obvious trend of decrease of $\beta$ with the increase of the quench temperature, starting with $\beta$=0.2 for T = 1136 K and ending with $\beta$=0.1 for T = 1217 K, the authors average $\beta$ over the temperature interval to obtain $\beta$=0.164, a value used further by other authors [13] as a reference one. Now this problem could be revisited with a new hypothesis for decrease of the range of step-step attraction $p$ (but still long-ranged, $p<n$=2) with increasing the quench temperature. In a parallel STM and theoretical study of the faceting kinetics on vicinal Si(113) [13] was found the same value of $\beta$ with the same error $\beta$=0.18±0.02. Our analysis of the digitized data confirms the exponent extracted from the experimental data and a more precise estimate of $\beta$ from the theoretical curve is $\beta$=0.203 (0.001). Thus one could make the conclusion that in the experiment $p$ is between from 0 to 2 most probably being around 1 (must be distinguishable from 2, the canonical value of $n$, in order to have well developed bunches, see equations 5 and 6).

In conclusion, we find a general result, Eq.7 for the dependence of the time-scaling exponent $\beta$ of the number of steps in the bunch $N$ on the range of the step-step attraction $p$ and this challenges further theoretical and experimental studies in the field.

**Acknowledgement:** The authors acknowledge financial support from IRC-CoSiM and MADARA Contracts with Bulgarian National Science Fund.

*e-mail: tonchev@ipc.bas.bg

1. J. Krug, V. Tonchev, S. Stoyanov and A. Pimpinelli, Phys. Rev. B **71**, 045412 (2005).
2. S. Baier, H. Ibach and M. Giesen, Surface Science **573**, (2004) 17.
3. H. Ibach and W. Schmickler, Surface Science **573**, (2004) 24.
4. D. Staneva, B. Ranguelov, and V. Tonchev, Nanosci. Nanotech. **10**, 11 (2010).
5. B. Ranguelov, V. Tonchev, and C. Misbah, in *Nanoscience and Nanotechnology*, E. Balabanova and I. Dragieva (eds.), Heron Press, Sofia, 2006, Vol. 6.
6. V. Tonchev, B. Ranguelov, H. Omi and A. Pimpinelli, EPJB **73**, (2010) 539.
7. H. Sakaguchi and N. Fujimoto, Phys. Rev. E **68,** 056103 (2003).
8. J.-K. Zuo, T. Zhang, J. F. Wendelken and D. M. Zehner, Phys. Rev. B **63**, 033404 (2001).
9. J. Tersoff, Y. H. Phang, Z. Zhang and M. G. Lagally, Phys. Rev. Lett. **75**, 14 (1995) 2730.
10. A. Pimpinelli, V. Tonchev, A. Videcoq and M. Vladimirova, Phys. Rev. Lett. **88**, 20 (2002) 206103.
11. S. Song, S. G. J. Mochrie and G. B. Stephenson, Phys. Rev. Lett. **74**, 26 (1995) 5240.
12. S. Song *et al.* , Surface Science **372**, (1997) 37.
13. K. Sudoh and H. Iwasaki, Phys. Rev. Lett. **87**, 21 (2001) 216103.

**Figure Captions**

Figure 1. Scaling of the 'minimal' step-step distance in the bunch with the model parameters, for MM2 model, data extracted from Monitoring Scheme II.

Figure 2. Time-scaling of the number of steps in the bunch $N$ (solid symbols) from LW2 and MM2(inset), $p = 0$ (MS – I). Also plotted are data for the time evolution of the terrace width (TW – hollow symbols) to show that it shares the same exponent with the bunch size $N$, in this case $\beta(p=0) \equiv \beta_0 = 1/5$ for LW2 (1/3 for MM2).

Figure 3. Time-scaling of the number of steps in the bunch $N$ from LW2 and MM2 (inset), $p = 0.5$ (MS – I). $\beta = 0.19$ for LW2 (0.3 for MM2).

Figure 4. Time-scaling of the number of steps in the bunch $N$ from LW2 and MM2 (inset), $n = 4$ (MS – I). For LW2 $p = 1$ and the best linear fit in log-log coordinates is $\beta = 0.18$ while for MM2 $p = 0.75, 1.5$ and the guiding-eye lines are plotted using equation 7 with $\beta = 0.28333, 0.23333$.

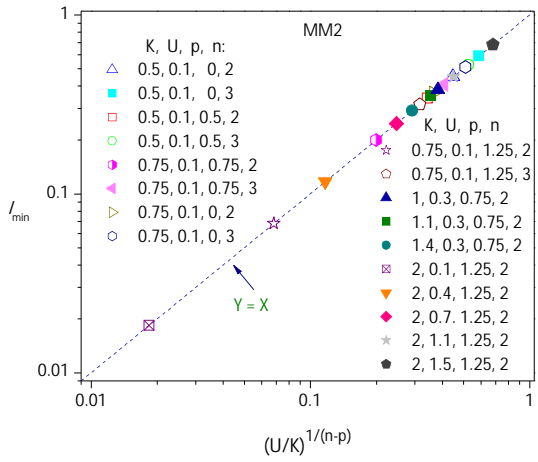

Figure 1

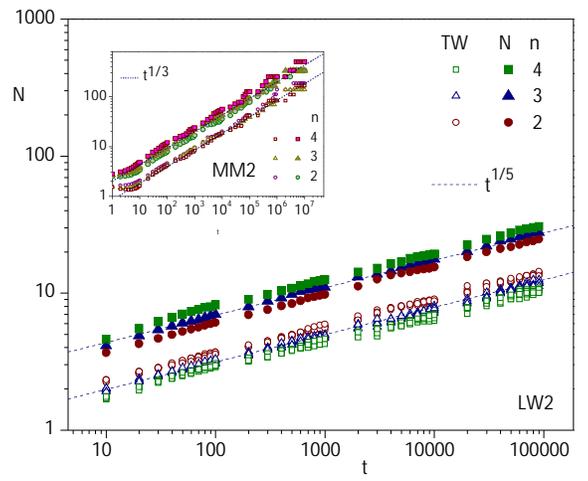

Figure 2

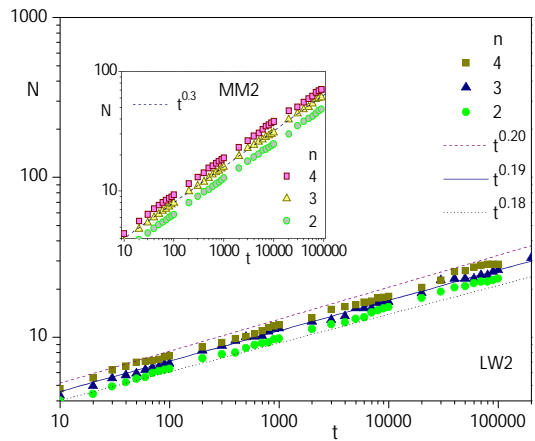

Figure 3

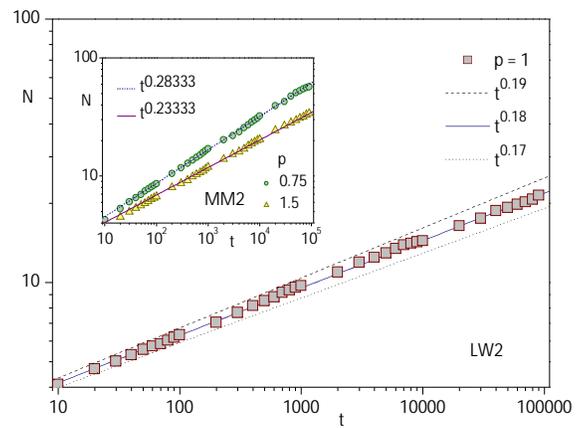

Figure 4

6